%% file: globecom_main.tex
\documentclass[conference]{IEEEtran}
\IEEEoverridecommandlockouts
\usepackage{dsfont}
\usepackage{cite}
\usepackage{amsmath,amssymb,amsfonts,bm}
\usepackage{graphicx}
\usepackage{textcomp}
\usepackage{xcolor}
\usepackage{mathtools}
\usepackage{placeins}

\usepackage{algorithm,algpseudocode,amsmath}
\def\BibTeX{{\rm B\kern-.05em{\sc i\kern-.025em b}\kern-.08em
    T\kern-.1667em\lower.7ex\hbox{E}\kern-.125emX}}
    
\begin{document}


\title{\huge{Collaborative Wideband Spectrum Sensing and Scheduling for Networked UAVs in UTM Systems }} 
    \author{
        \IEEEauthorblockN{Sravan Reddy Chintareddy\IEEEauthorrefmark{1}, Keenan Roach\IEEEauthorrefmark{2}, Kenny Cheung\IEEEauthorrefmark{2}, Morteza Hashemi\IEEEauthorrefmark{1}}
        \IEEEauthorblockA{
        \IEEEauthorrefmark{1}Department of Electrical Engineering and Computer Science, University of Kansas \\
    \IEEEauthorrefmark{2}Universities Space Research Association (USRA)
        }
    }


\maketitle

\begin{abstract}
\input{abstract}
\end{abstract}


\input{introduction_final}

\input{background_final}

\input{problem_final}
\input{algorithm}
\input{results_final}
\input{conclusion}
\vspace{-.1cm}
\section{Acknowledgement}
The material is based upon work supported by NASA under award No(s) 80NSSC20M0261, and NSF grants CNS-1948511 and CNS-1955561. Any opinions, findings, and conclusions or recommendations expressed in this material are those of the author(s) and do not necessarily reflect the views of the National Aeronautics and Space Administration (NASA) and the National Science Foundation (NSF).

\vspace{-.1cm}
\bibliographystyle{IEEEtran}
\bibliography{globecom_main}

\end{document}

%% file: abstract.tex
In this paper, we propose a
data-driven framework for collaborative wideband spectrum sensing and
scheduling for networked unmanned aerial vehicles (UAVs), which act as the secondary users to opportunistically utilize detected spectrum holes. To this end, we propose a multi-class classification problem for wideband spectrum sensing to detect vacant spectrum spots based on collected I/Q samples. To enhance the accuracy of the spectrum sensing module, the outputs from the multi-class classification by each individual UAV are fused at a server in the unmanned aircraft system traffic management (UTM) ecosystem. In the spectrum scheduling phase, we leverage reinforcement learning (RL) solutions to dynamically allocate the detected spectrum holes to the secondary users (i.e., UAVs). To evaluate the proposed methods, we establish a comprehensive simulation framework that generates a near-realistic synthetic dataset using MATLAB LTE toolbox by incorporating base-station~(BS) locations in a chosen area of interest, performing ray-tracing, and emulating the primary users channel usage in terms of I/Q samples. This evaluation methodology provides a flexible framework  to generate \emph{large spectrum datasets} that could be used for developing ML/AI-based spectrum management solutions for aerial devices.  

%% file: introduction_final.tex
\vspace{-.2cm}
\section{Introduction}
\label{sec:intro}
Unmanned aerial vehicles (UAVs) have attracted significant interests from communications and networking, robotics, and control societies for exploring novel applications such as on-demand connectivity, search-and-rescue operations, situational awareness, to name a few~\cite{menouar2017uav}. 
In order to truly unleash the potentials of UAVs, real-world and commercial deployments will most likely be in the form of beyond visual line-of-sight (BVLOS) scenarios, which in turn provide easier access to remote or hazardous areas, less human intervention, and reduced cost of operation~\cite{li2019beyond}. 
For safe operations of multiple UAVs under BVLOS conditions, the NASA and FAA are in the process of defining the Unmanned Aircraft System Traffic Management (UTM) systems~\cite{kopardekar2016unmanned}.

Existing terrestrial mobile networks (e.g., 4G LTE and the upcoming 5G-and-beyond) provide significant wireless coverage with relatively low latency, high throughput, and low cost. This, in turn, will make the cellular network a good  candidate for the operation of UAVs in BVLOS scenarios~\cite{abdalla2021communications}. 
However, the proliferation of new wireless services and the demand for higher cellular data rates have significantly exacerbated the spectrum crunch that cellular providers are already experiencing. Therefore, developing  dynamic spectrum management services to
sense, assign, and monitor spectrum usage within the UTM architecture is of utmost importance in order to enable advanced UAV use cases in BVLOS, such as urban air mobility (UAM) and advanced air mobility (AAM)~\cite{rimjha2021urban}. 

There exists a multitude of prior works on spectrum management frameworks for ground users~\cite{zhang20196g,ahmad20205g,uvaydov2021deepsense,cui2019multi,li2020deep,nguyen2018deep}. For instance, the authors in ~\cite{li2020deep,uvaydov2021deepsense,cui2019multi} propose deep learning based wideband spectrum sensing to dynamically detect ``spectrum holes''. Furthermore, several studies leverage reinforcement learning (RL) techniques for spectrum sharing, assuming that the spectrum sensing results are readily available~\cite{nguyen2018deep}. While these data-driven spectrum management frameworks for ground users are available, they are not directly applicable for UTM-enabled UAV operations, due to several factors, such as widely different wireless channel models and the overall system architecture~\cite{uvaydov2021deepsense,nguyen2018deep}. 
In the context of UAV spectrum systems, the authors in~\cite{kakar2017waveform,shang2020spectrum} have developed efficient spectrum sharing policies for UAV communications to enhance the spectral efficiency (SE). However, the majority of the existing spatial spectral sensing (SSS) models do not consider the spectrum usage pattern of users under realistic scenarios (e.g., ignoring the I/Q level samples), and/or they consider only a single primary user (PU) or secondary user (SU). Moreover, the problem of joint multi-channel wideband spectrum sensing and scheduling among several SU's has not been fully investigated. 

To address these gaps, in this paper, we propose a data-driven model for joint wideband spectrum sensing and scheduling across several UAVs, which act as secondary users to opportunistically utilize detected spectrum holes.  Our proposed system model presents a unified framework that is compatible with the UTM deployment models with centralized controlling and monitoring entities (e.g., UAS service suppliers). To make our development more concrete and grounded, the problem of joint spectrum sensing and sharing is formulated as an energy efficiency (EE) maximization in a wideband multi-UAV network scenario. Then, we transform the EE optimization problem into a Markov Decision Process (MDP) to maximize the overall throughput of the SUs. To enable spectrum sensing, we develop a multi-class classification framework to identify spectrum holes based on observed I/Q samples. To enhance the accuracy of the spectrum sensing module, the outputs from the multi-class classification by each individual UAV are fused at the UTM server. In the spectrum scheduling phase, we develop and implement several RL algorithms, including the standard Q-learning methods to dynamically allocate underutilized spectrum sub-channels to multiple UAVs. We further investigate the performance of the ``vanilla'' deep Q-Network (DQN) and its variations, including double DQN (DDQN) and DDQN with soft-update.

 Furthermore, one of the primary challenges of using machine learning (ML) methods for spectrum management is the need for large amounts of training data. The lack of available spectral data in many cases is a significant obstacle, especially for UAV networks that introduce additional complexity for large-scale experimental data collection. To address this gap, and evaluate the proposed methods, we develop a comprehensive framework for spectrum dataset generation, which accurately models LTE waveform generation and propagation in any environment of interest for UTM-enabled UAV applications. This platform enables modeling cooperative spectrum sensing and sharing for wideband multi-UAV network scenarios, and can be used for scalable generation of large spectrum datasets within an area of interest.


The rest of this paper is organized as follows. In Section~\ref{sec:background}, we review related works.  In Section~\ref{sec:problem}, we discuss our system model, followed by the problem formulation for joint spectrum sensing and access in Section~\ref{sec:problem2}.  Section~\ref{sec:dataset} describes our methodology to generate synthetic spectrum dataset, and Section~\ref{sec:results} includes our numerical results. Section~\ref{sec:conclusion} concludes the paper.

%% file: background_final.tex
\section{Related Works}
\label{sec:background}
\noindent
\textbf{Works on Spectrum Sensing for UAVs.} 
The authors in~\cite{shang2019spatial} address spectrum access and interference management by utilizing SSS for ground based device-to-device (D2D) communications~\cite{chen2016spatial,chen2018qos}. Furthermore, the authors in~\cite{shang2020spectrum} extend the usage of SSS to UAVs. The UAVs perform SSS to obtain the received signal strength and compare it with a threshold to identify the spectrum occupancy of a particular uplink channel, similar to energy detection methods. The performance of the SSS is then characterized using the probability of spatial false alarm and probability of spatial miss detection. The SSS methods are not directly applicable to wideband sensing to detect multiple spectrum holes simultaneously by all UAVs. 

In addition to the SSS methods, data-driven deep learning (DL) methods for spectrum sensing have been considered in several works~\cite{liu2019deep,chew2020spectrum}. However, multi-channel wideband sensing using DL methods have not been explored by the above studies. To address this, the authors in~\cite{uvaydov2021deepsense} developed a DL-based fast wideband spectrum sensing. The DL model is based on convolutional neural networks (CNN) and accepts raw I/Q signals and predicts the spectrum holes. This work considers a single PU, a single SU only, and the channel between them is modeled as a Rayleigh fading channel.

In this paper, we consider dataset generation for downlink LTE with multiple transmitters~(base stations) and receivers~(UAV), thereby incorporating interference caused by neighboring cells. Moreover, we use real deployment LTE BS locations for transmitters and employ ray-tracing to effectively model the dynamic UAV environment. Additionally, we consider collaborative spectrum sensing and aggregate the spectrum hole results. We show that spectrum fusion improves the reliability of overall spectrum hole detection.


\noindent
\textbf{Works on Spectrum Sharing methods.} Spectrum sharing solutions for ground based communications has been extensively investigated (e.g.,~\cite{zhang20196g,ahmad20205g,sharmila2019spectrum,slamnik2020sharing}). The authors in~\cite{li2020deep,naparstek2017deep} propose the use of RL for dynamic spectrum access in multi-channel wireless networks. Moreover,~\cite{nguyen2018deep,bokobza2023deep,wang2018deep} proposes the use of DQN where in each time slot a single SU decides to stay idle or transmit using one of the sub-channels in a multi-channel environment. While these studies have provided significant insights, they consider one SU only.

Compared with the existing works in this area, the key contributions of this paper are as follows: \textbf{(i)} we develop a \emph{joint} spectrum sensing and access framework using raw LTE I/Q data, \textbf{(ii)} the spectrum sensing module identifies multiple spectrum holes in a wideband multi-channel setting, \textbf{(iii)} we use RL-based techniques (i.e., DQN and its variations) to allocate the identified spectrum holes to multiple UAVs, and \textbf{(iv)} the developed solution is based on realistic channel modeling between several LTE BSs and UAVs.

%% file: problem_final.tex
\section{System model}
\label{sec:problem}
\textbf{Network Model and Communication Protocol.} We consider a set of UAVs denoted by $\mathcal{K}$ ($|\mathcal{K}| = K$) where each UAV is capable of performing wideband sensing over $M$ orthogonal primary spectrum resources (sub-channels) independently. Due to the highly dynamic environment in which UAVs operate, it may not be feasible for all the UAVs to observe every vacant sub-channel. Therefore, we leverage a collaborative spectrum sensing by the UAVs, and perform spectrum fusion at the UTM server to increase the reliability of spectrum hole detection. Identified spectrum holes are allocated to the UAVs. Therefore, the overall system model is divided into two major components: (i) collaborative spectrum sensing and fusion policies, (ii) spectrum allocation and access policies. 

To coordinate the spectrum sensing, fusion, and access steps,
we assume that each  time slot is divided into four consecutive sub-slots:  UAV resource request~($t_{req}$), spectrum sensing~($t_s$), broadcasting to server~($t_b$), and channel access~($t_a$). 
Specifically, at the beginning of each time slot, the UAVs that require PU resources request the server for resource allocation. 
In the subsequent sub-slot of sensing~($t_s$), the UAVs perform spectrum sensing and broadcasts the sensed channel information in the following sub-slot~($t_b$). The server then applies fusion rules and assigns spectrum holes to the requesting UAVs. The UAVs transmit on the allocated spectrum holes in the access sub-slot~($t_a$).


\textbf{Collaborative Spectrum Sensing and Fusion Policies.}
\label{sec: sensing module}
Each individual UAV captures the raw I/Q samples from over the air received signals and predict the availability of spectrum holes across $M$ sub-channels. We assume that there is an associated spectrum sensing cost for each UAV \emph{k} involved in sensing at time slot \emph{t}.   The spectrum sensing cost is the energy consumed for sensing the spectrum and is proportional to the voltage $V_{CC}$ of the receiver, the system bandwidth $B$, and the duration allotted for sensing~($t_s$)~\cite{zhang2011mili}. Therefore, it is defined as \small $SC_{km}(t) = t_sV_{CC}^2B_{m}$. \normalsize Upon the completion of sensing phase, the UAV $k$
has a predicted spectrum occupancy vector $\textbf{h}_k(t) = [h_{k, 1}(t), ..., h_{k,M}(t)]$ such that $h_{k, m}(t) = 0$ if the $m$-th sub-channel is detected vacant at time $t$, and $h_{k, m}(t) = 1$ otherwise. This problem can be considered as a multi-class classification problem, and we leverage deep neural network (DNN) at each UAV to identify the spectrum holes and outputs the prediction vector $\textbf{h}_k(t)$. 





The UTM server receives multiple copies of spectrum holes detected by individual UAVs and applies fusion rules that results in aggregated spectrum holes. In this paper, we use $n$-out-of-$N$ fusion rule that is defined as follows: 

\vspace{-.2cm}
\small
\begin{equation}
\label{eqn:fusion}
    f_m(t) = \begin{dcases}
    0,& \text{if } \sum_{k \in \mathcal{K}} \mathds{1} \{h_{k, m}(t)=0\}\geq n;\\
    1,              & \text{otherwise},  
\end{dcases}
\end{equation}
\normalsize
where $\mathds{1}\{.\}$ is an indicator function. In this case,  $\textbf{f}(t)=[f_1(t), ..., f_M(t)]$ is the fused prediction of all the $M$ sub-channels at the UTM server. 
Note that when $n=1$, the $n$-out-of-$N$ rule is equivalent to the ``OR'' rule, and $n=N$ is the same as the ``AND'' rule. 


 \textbf{Spectrum Allocation and Access Policies.}
\label{sec: access module}
Based on the aggregated fusion result, the UTM server then allocates sub-channels to the requesting UAVs. The UAVs then transmit data on the sub-channels allocated to them by the server in the next time step. 
The transmission energy consumption is denoted by \small $AC_{km}(t)$ \normalsize. The access cost is the energy consumed for data transmission and is defined as \small $AC_{km}(t) = t_{a}P_{tx}$, \normalsize 
where, $P_{tx}$ is the transmit power and $t_{a}$ is the time allotted to transmission. Furthermore, the transmission utility is the amount of data transmitted on the allocated sub-channel and is defined as follows:

\vspace{-.2cm}
\small
\begin{equation}
    R_{km}(t)= t_a B_m \log_2\left(1+\text{SINR}_{k,m}(t)\right),
\end{equation}
\normalsize
where $B_m$ is the sub-channel bandwidth allocated for data transmission and $\text{SINR}_{km}$ is the signal-to-interference-plus-noise ratio observed on the link between UAV and its receiver over sub-channel $m$. 

\begin{figure}[!t]
    \centering
    \includegraphics[width=\linewidth]{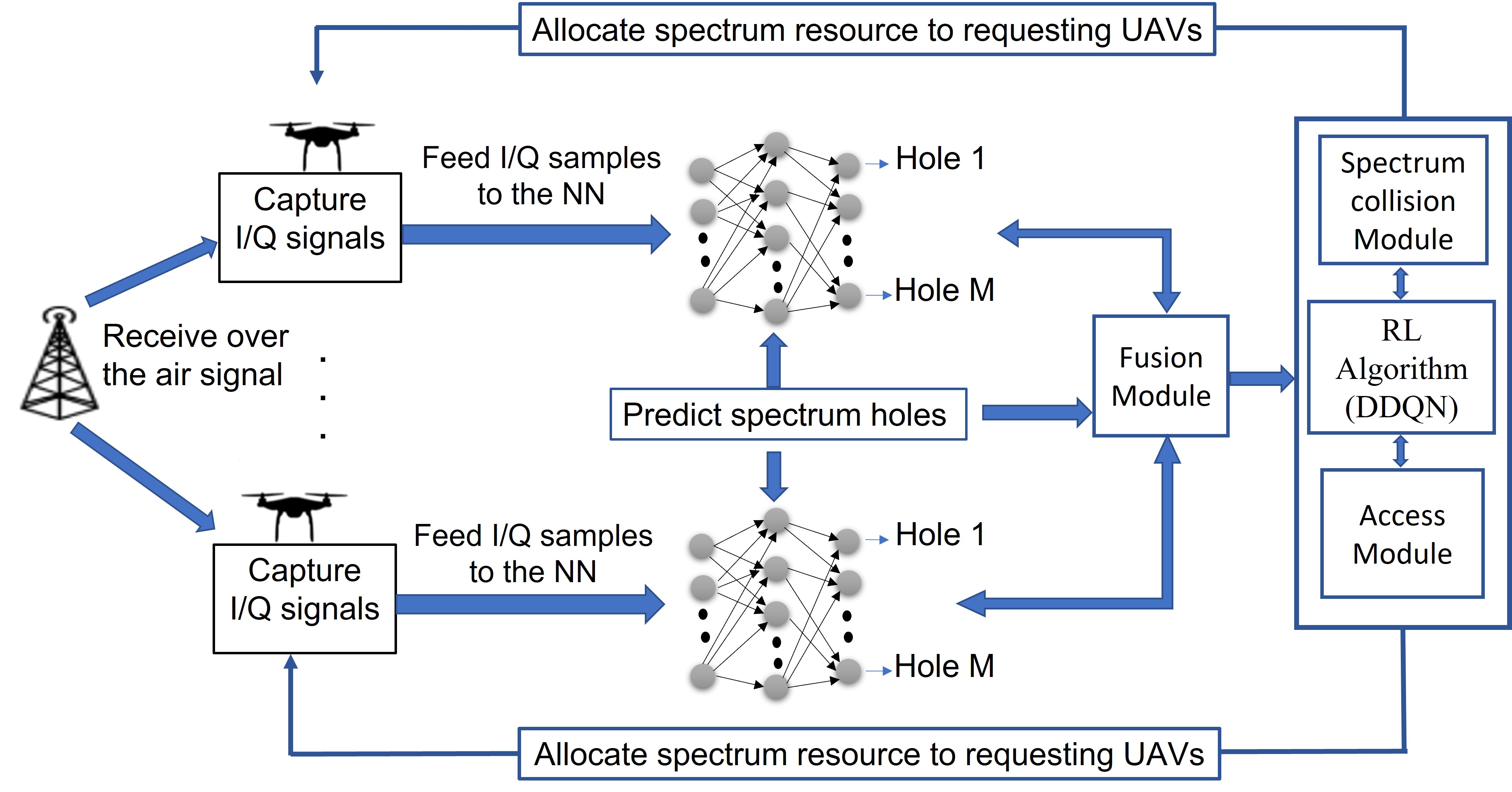}
     \vspace{-5mm}
    \caption{Proposed system model.}
    \label{fig:sysmodel} 
    \vspace{-6mm}
\end{figure}



Since it is delayed a transmission, the UAVs transmit on the spectrum holes in the current step that are detected vacant in the previous time step. Spectrum collision occurs when the previously detected spectrum holes are no longer available at the current time step. We assume that the true state of sub-channel $m$ is denoted by $\bar{f}_m(t)$. To capture this, we define the spectrum access collision indicator $r_{km}(t)$ as follows: 

\vspace{-.1cm}
\small
\begin{equation}
    r_{km}(t) = \begin{dcases}
        1, & \text{if } \bar{f}_{m}(t)=0~ \text{and}~  f_{m}(t-1) =0; \\
        -1, & \text{if } \bar{f}_{m} (t)\neq0~ \text{and} ~f_{m}(t - 1) =0; \\
        0, & \text{otherwise}. 
    \end{dcases}
\end{equation}
\normalsize


\section{Joint Spectrum Sensing and Access}
\label{sec:problem2}
Given the presented model, we cast the problem of joint spectrum sensing and access as an energy efficiency optimization for the UAVs. The overall system model of  collaborative spectrum sensing and access is shown in Fig. \ref{fig:sysmodel} with the overall algorithm described in Algorithm~\ref{alg:algo2}. 


\textbf{Energy Efficiency Optimization.}
Let $y_{km}(t)=1$ if UAV $k$ is scheduled to use sub-channel $m$ at time $t$, and $y_{km}(t)=0$ otherwise. Given that the spectrum holes are allocated to the requesting SUs based on the sub-channel availability we incorporate sensing and access cost to maximize the overall energy efficiency (EE) of the system. We formulate EE problem as an optimization problem as follows:

\vspace{-.4cm}
\begin{equation}
\label{eqn:op3} 
\begin{cases}
\mathop{\mathrm{max}}\limits_{\{y_ {km}(t)\}} & \mathrm{E} \big\{ \sum\limits_{t,k,m} \frac{\ y_ {km}(t) \   r_ {km}(t) \ R_ {km}(t)} {y_ {km}(t) AC_ {km}(t) + SC_{km}(t) }  \big\}  \\
\text{subject to:} &  \sum_{m} \ y_ {km}(t) \ \leq 1 , \ \forall \ k=1,2,3, \dots K , \\
&  \sum_{k}  \ y_ {km}(t) \ \leq 1 , \ \forall \ m=1,2,3, \dots M , \\
&  \sum_{k,m} \ y_ {km}(t) \ \leq \ M -|\textbf{f}(t)| , \ \\
&   y_ {km} (t) \in \{0,1\},  
\end{cases}
\vspace{-0.5mm}
\end{equation}
where $R_{km}(t)$, $SC_{km}(t)$, and $AC_{km}(t)$ are the amount of data transmitted, the sensing cost, and transmission cost by the SU~$k$ on sub-band $m$ simultaneously. The constraints guarantee that each UAV is scheduled to use at most one sub-channel, while the total number of scheduled UAVs is at most equal to the number of detected spectrum holes at time $t$, which is $M-|\textbf{f}(t)|$.

This optimization problem  is a fractional integer programming problem, which is NP-hard in general. If we consider maximizing the numerator alone, which is the total utility of the UAVs over all sub-channels, the problem will become an integer programming problem. In this case, the utility would depend on the spectrum usage pattern by the PUs, which is captured by $r_{km}(t)$ as well as the channel condition between the BSs and UAVs that determine the amounts of transmitted data $R_{km}(t)$. To tackle this utility optimization problem,  we model the channel occupancy $\bar{f}_m(t)$ as a Markov process, which enables us to use an Markov decision process (MDP) formulation to solve this problem~\cite{sutton2018reinforcement} and develop a dynamic spectrum allocation policy to the SUs.  
\newcounter{phase}[algorithm]
\newlength{\phaserulewidth}
\newcommand{\setphaserulewidth}{\setlength{\phaserulewidth}}
\newcommand{\phase}[1]{%
  \Statex\leavevmode\llap{\rule{\dimexpr\labelwidth+\labelsep}{\phaserulewidth}}\rule{\linewidth}{\phaserulewidth}
  \Statex\strut\refstepcounter{phase}\textit{Phase~\thephase~--~#1}
  \vspace{-0.5ex}\Statex\leavevmode\llap{\rule{\dimexpr\labelwidth+\labelsep}{\phaserulewidth}}\rule{\linewidth}{\phaserulewidth}}
\makeatother

\setphaserulewidth{.3pt}

\begin{algorithm}[!hb]
  \caption{Collaborative Spectrum Sensing and Access}
  \label{alg:algo2}  
  \begin{algorithmic}[1]     
    \phase{Spectrum Sensing and Broadcasting }
    \For{each UAV in~$\mathcal{K}$}
        \State Capture $N$ I/Q samples from over the air signal, \hspace*{1.2em}  where, \textbf{X} $\in \mathbb{C}^{N\times 2}$.
        \State Feed I/Q samples to the pre trained ML model that \hspace*{1.2em} predicts the spectrum holes s.t $f:\textbf{X} \rightarrow \textbf{h}$.
        \State Broadcast the individual spectrum hole observations  \hspace*{2.2em} \textbf{h} $\in \{0,1\} ^{1 \times m}$ to the fusion center.      
        \EndFor \vspace{-1mm}
    \phase{Spectrum Fusion and Access}
     \State Apply fusion rule in Eq.~\eqref{eqn:fusion}  to predict spectrum holes $\textbf{f}(t)$.
    \State Allocate a single spectrum hole to each requesting UAV using pre trained RL algorithm, $y_{km}(t)$, s.t. the constraints in Eq.~\eqref{eqn:op3} are satisfied.
    \State UAVs transmit on the sub-channel allocated in the previous allocated time slot
    
    \State Given the spectrum allocation $y_{km}(t)$ and spectrum access collision indicator $r_{km}(t)$, the total utility U(t) can be computed using Eq.~\eqref{eqn:op3}.
\end{algorithmic}

\end{algorithm}

\textbf{Dynamic Spectrum Allocation Using RL.}
\label{sec:DSA}
As we assume there exists $M$ sub-channels in the system, each sub-channel can be modeled as an independent two-state Markov chain. The transition probability function \textbf{P}, can then be viewed as a set of transition probability matrices \{$\textbf{P}_i$\} for each sub-channel that capture the randomness in the assumed multi-user multi-channel environment. Hence, we can formulate the total utility of the SUs into a traditional MDP which is governed by the tuple ($\mathcal{S}$, $\mathcal{A}$, \{$\textbf{P}_i$\}, U, $\gamma$), consisting of the set of states $\mathcal{S}$, set of actions $\mathcal{A}$, a transition probability function \{$\textbf{P}_i$\}, a reward function $U$, and a discount factor $\gamma$. To solve an MDP using RL, an agent learns to make decisions in an uncertain environment by maximizing a cumulative reward over a sequence of actions. Specifically, the agent interacts with an environment by taking actions that transition the system from one state to another, and the agent receives a reward that is commensurate with the merit of the action. The discount factor determines the relative importance of immediate and future rewards.  One of the most popular RL methods is Q-learning~\cite{sutton2018reinforcement}.
The classical Q-learning is table-based, i.e. the values of the Q-function are stored in a table of size $|\mathcal{S}|$×$|\mathcal{A}|$. However, when the size of the state and action spaces get large, the complexity of tabular Q-learning becomes cumbersome. For example, with $M = 16$ sub-channels, the Q-table will be of size $65,537 \times 17$.




\textbf{DDQN-Based Spectrum Allocation.} To address the complexity issue, we adapt the deep Q-learning approach in~\cite{van2016deep} to approximate the Q-function by a neural network $Q_{\bm{\theta}}$ called Double Deep Q-Network (DDQN) and train its weights $\bm{\theta}$ using experience replay. As the name suggest, we have two networks when using DDQN where, $Q_{\bm{\theta}}$ is called the primary network and $Q'_{\bm{\theta}}$ is called the target network and the weights of the target network are updated periodically. In the original DDQN, the weights of target network are directly copied from the primary network every few episodes. In DDQN-soft, the target networks are updated using polyak averaging to smoothly update the weights (``soft-update'')~\cite{van2016deep}.

\begin{figure}[!t]
 \centering
    \includegraphics[width=0.48\textwidth]{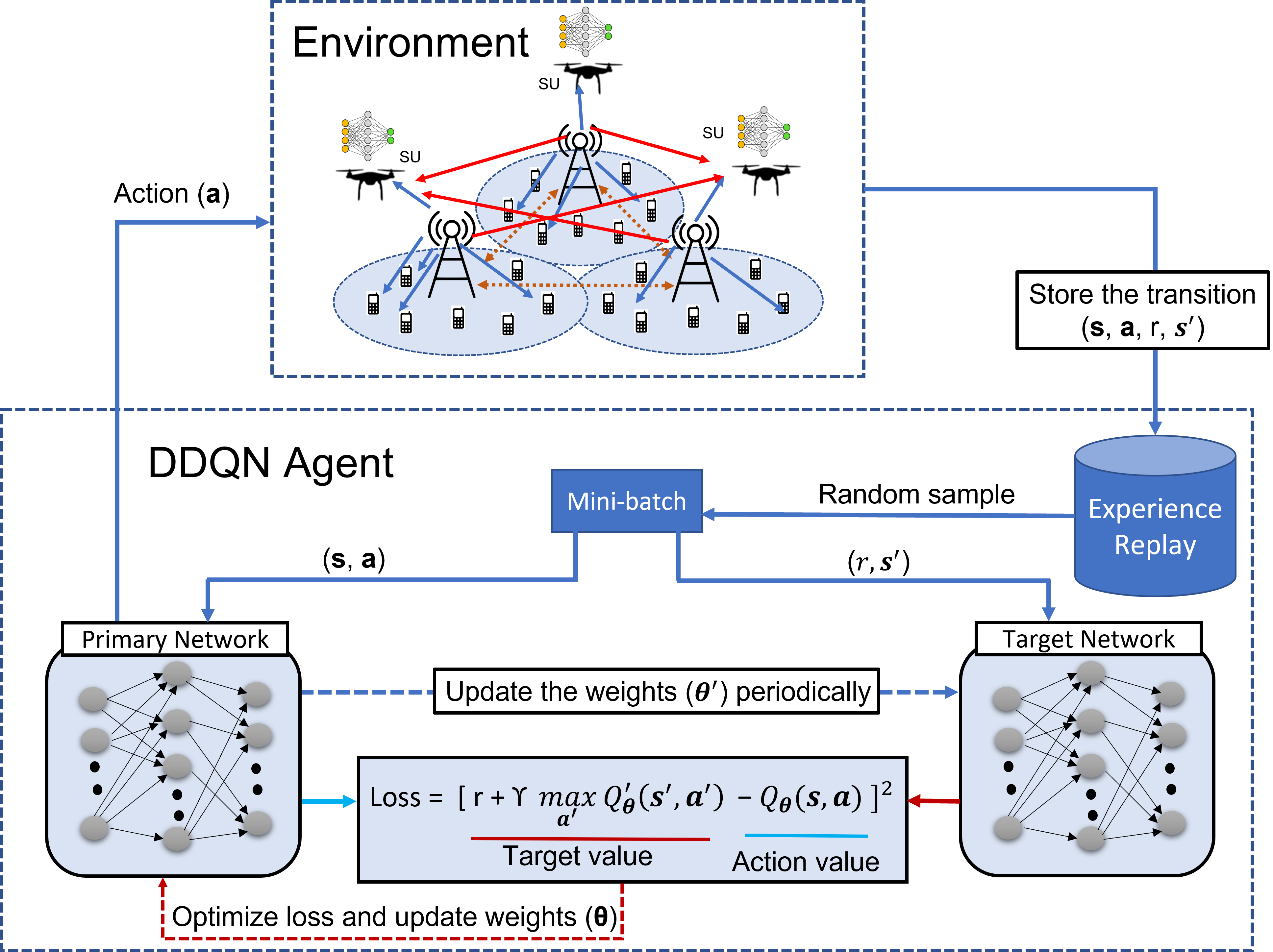}
     \vspace{-3mm}
    \caption{DDQN for spectrum access.}
    \label{fig:ddqndsa}  
    \vspace{-5mm}
\end{figure}
The input to the DDQN agent is a state \textbf{s} of size $1 \times M$ . The output of the network is a vector of size $1 \times (M + 1)$ that contains the values of the Q-function with respect to state \textbf{s} and each of the $M + 1$ actions. In all the hidden layers, we use the rectified linear unit~(ReLU) as an activation function. Given the neural networks input-output dimensions, the overall DDQN architecture and its interaction with the environment is shown in Fig.~\ref{fig:ddqndsa}. As shown in Fig.~\ref{fig:ddqndsa}, the major components are primary network, target network, experience replay and the interaction with the environment to pick an action.

To train the DDQN agent, the experiences are initially stored in the memory using $\epsilon -$greedy policy i.e., for a state $s_t$, an action $a_t$ is taken randomly with probability $\epsilon_t$ or taken greedily with probability $1- \epsilon_t$ from the current state of the DDQN network. Then, when we have sufficient samples in the memory a mini-batch of $B$ experiences $\{(\textbf{s}_i,\textbf{a}_i,r_i,\textbf{s}_i ')\}_i \in \mathcal{B}_t$ is randomly sampled from the memory for every time step \emph{t} to train the neural networks. Here, $\mathcal{B}_t$ is the set of experiences currently available in the memory. Based on the mini-batch selected, we compute and update the weights $\bm{\theta}$ of the primary network $Q_{\bm{\theta}}$ that minimize the loss function $L_t({\bm{\theta}})$. Fig.~\ref{fig:ddqndsa} captures the overall DDQN architecture and the agents interaction with the environment~\cite{sutton2018reinforcement,van2016deep}. 

%% file: results_final.tex
\vspace{-.2cm}
\section{Raw I/Q Dataset Generation}
\label{sec:dataset}
In the previous section, we described the role of DDQN in allocating spectrum holes to the SUs such that the overall utility is maximized. However, each SU has to first send in their spectrum hole detection results based on observed I/Q samples. 
Implementing a data-driven ML model for such wideband sensing demands large amounts of raw I/Q data. 
While it is desirable to capture over-the-air raw I/Q signals using actual hardware, it is challenging to accomplish this goal due to the intricate nature of flying multiple UAVs within a specific environment for collaborative sensing implementation.
Hence, we resort to generating synthetic dataset that accurately resembles collecting dataset via experimentation. 
To this end, we use MATLAB's LTE toolbox as outlined in~\cite{uvaydov2021deepsense} and extend the dataset generation to incorporate UAV specifics. In particular, we generate the dataset by incorporating the LTE base-stations and UAVs locations, as well as 3D environment including buildings and vegetation for performing ray-tracing.

\vspace{-1mm}
\begin{figure}[!htb]
\centering    \includegraphics[width=0.35\textwidth]{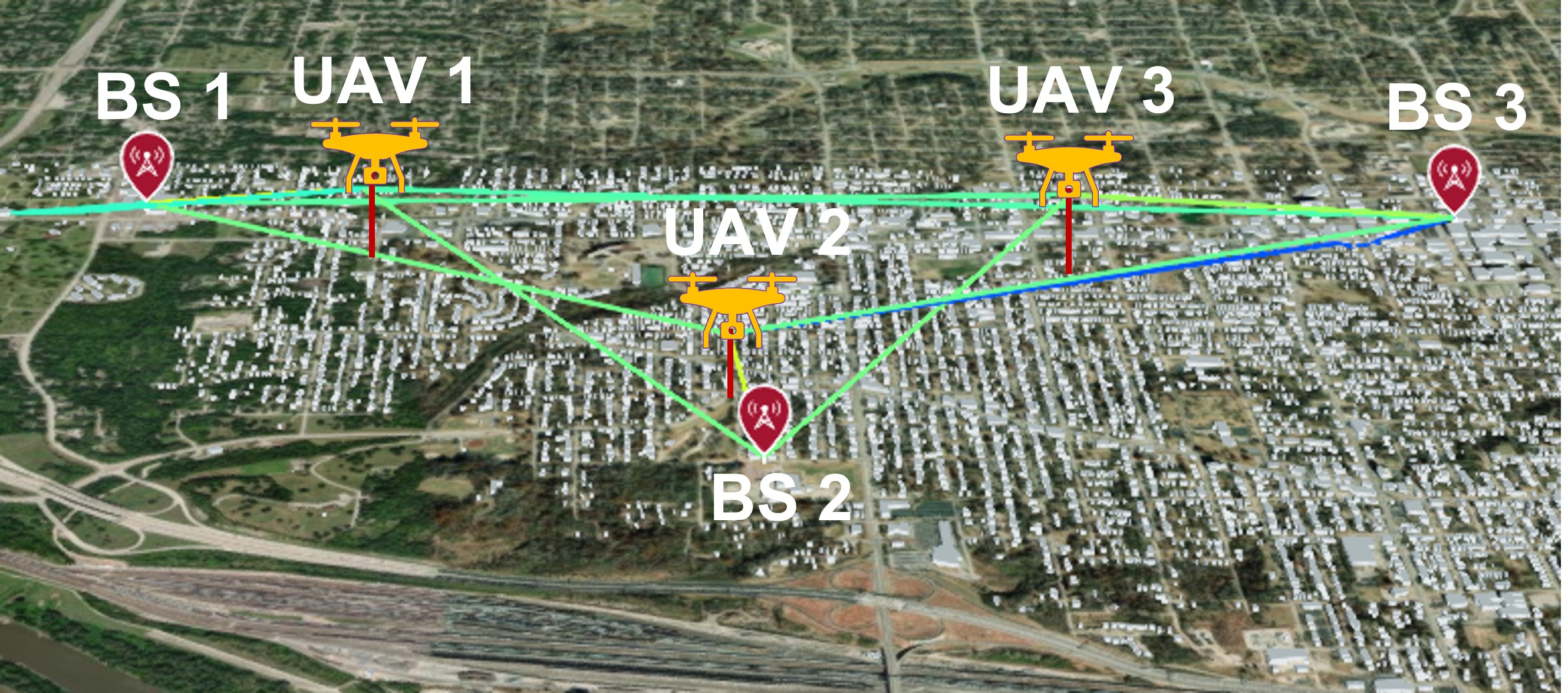}
    \vspace{-3mm}
    \caption{Simulation setup for dataset generation }    
    \label{fig:simsetup}
     \vspace{-4mm}
\end{figure}
\textbf{Dataset Generation Methodology.} As shown in Fig.~\ref{fig:simsetup}, we assume three neighboring cells.  In order to perform ray-tracing experiments, the maps required are downloaded from OpenStreetMap~\cite{OSM}.


The simulation area we considered is of $3$ km $\times$ $3$ km wide with buildings and vegetation. The location of the base stations are obtained from cellmapper~\cite{cellmap}, an open crowd sourced cellular tower and coverage mapping service. We define base-stations as the transmitter sites and UAV locations as the receiver sites. 
Furthermore, we consider three UAVs and three base stations in the region of interest, and use MATLAB's ray-tracer to find the channel between UAV and base-station locations. Note that this scenario can be easily extended to any numbers of LTE cells and UAVs. It is important to emphasize that we consider UAVs hovering in one location. Nevertheless, by executing the ray-tracing engine multiple times for different locations, we can effectively simulate a flight trajectory.

In the next step, we use MATLAB's LTE Toolbox to generate LTE-M waveform. The entire cell bandwidth of $10$ MHz ($50$ resource blocks) is assumed to be split into $16$ sub-channels each of size three resource blocks. In general, the base station can allocate a single sub-channel or multiple sub-channels to a PU to transfer user specific data on the downlink shared channel and also can use multiple access techniques for transmitting data to different PUs. However, when we generate the dataset we assume that base station partitions the bandwidth into separate sub-channels. While generating the downlink waveform of the base station we do not generate any UE specific reference  signals. Also, we do not corrupt the broadcast channels with user-specific data. We find the appropriate indices and embed the data samples into the downlink shared channel. As we assumed we partition the cell bandwidth into $16$ sub-channels, considering each combination as a label, the base station can generate $2^{16}$ labels, ranging from no sub-channel allocation to a fully busy cell site. 
Since we assumed all SUs are capable of wideband sensing, each SU samples the RF signal and stores I/Q samples. We vary the noise variance such that the effective SINR varies from $-10~\text{dB}$ to $20~\text{dB}$ in steps of $10~\text{dB}$. Here, the interference caters for the additional received signals from the neighboring base stations which are modeled using the ray-tracing setup as shown in Fig.~\ref{fig:simsetup}.

\begin{figure}[!htb]
\minipage{0.48\linewidth}
\centering
  \includegraphics[width=\linewidth]{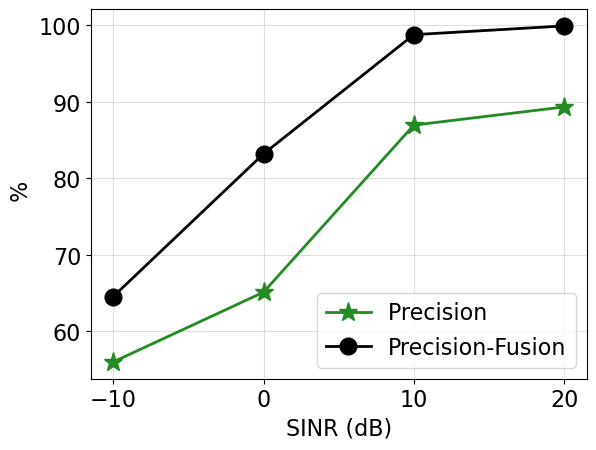}
  \vspace{-7.5mm}
  \caption{\textbf{Precision} performance at location 3.}
  \label{fig:prec}
\endminipage\hfill
\minipage{0.48\linewidth}
\centering  \includegraphics[width=\linewidth]{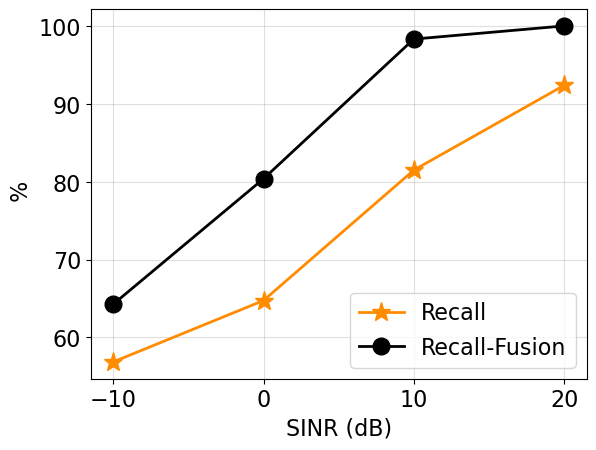}
  \vspace{-7.5mm}
  \caption{\textbf{Recall} performance at location 3.}
  \label{fig:rec}
\endminipage\hfill
 \vspace{-6mm}
\end{figure}

\vspace{-.3cm}
\section{Numerical Results}
\label{sec:results}
\textbf{Collaborative Spectrum Sensing Results.}
As mentioned earlier, identifying spectrum holes falls into the realm of classical multi-class classification problem. We consider Precision, Recall, and F1-score as the metrics to assess the performance of such classifier. The general definitions of Precision and Recall are as follows: 

\vspace{-.2cm}
\small
\begin{equation}
     \text{Precision} = \frac{\text{TP}}{ \text{TP} + \text{FP}} ,\quad  \text{Recall} = \frac{\text{TP}}{ \text{TP} + \text{FN}},
\end{equation}   
\normalsize
where TP, FN, FP accounts for the number of true positives, false negatives, and  false positives, respectively. 
To concretely capture the performance of spectrum sensing across $16$ sub-channels, we compute the micro-averages for Precision and Recall~\cite{grandini2020metrics}.

We use $70\%$ of the samples generated to train the DNN while the rest of samples are used for testing and validation purposes. We observed that the performance metrics improve as the SINR improves. Specifically, for UAV locations $1$ and $2$, the spectrum sensing performance metrics are greater than $90\%$ for the SINR values above $10~\text{dB}$. However, for UAV location 3, the overall performance metrics are worse than locations 1 and 2. As shown in Fig.~\ref{fig:prec} and \ref{fig:rec},  the performance metrics are about $90\%$ at $20~\text{dB}$ SINR. This is due to weaker received signal strength in location 3, leading to more noisy I/Q samples. Hence, the ML model was not able to predict the spectrum holes accurately, thereby making the rationale for exploring collaborative sensing more apparent.
From the spectrum fusion results (described in Eq.~\eqref{eqn:fusion}) shown in Fig.~\ref{fig:prec} and \ref{fig:rec}, we note that incorporating predictions from UAV locations 1 and 2 significantly 
improved the spectrum prediction performance. 

\begin{figure}[!t]
\minipage{0.48\linewidth}
\centering
 \includegraphics[width=\linewidth]{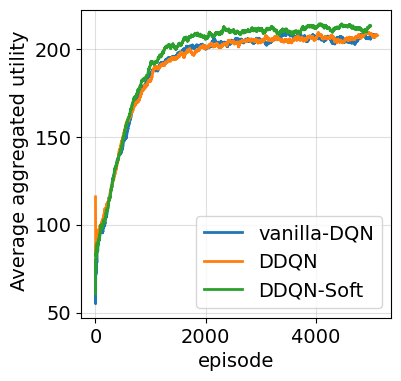}
    \vspace{-8mm}
    \caption{Training results for allocating spectrum holes to \textbf{one UAV}.}
    \label{fig:dqn-1UAV}
\endminipage
\vspace{-7mm}
\minipage{0.48\linewidth}
\centering
  \includegraphics[width=\linewidth]{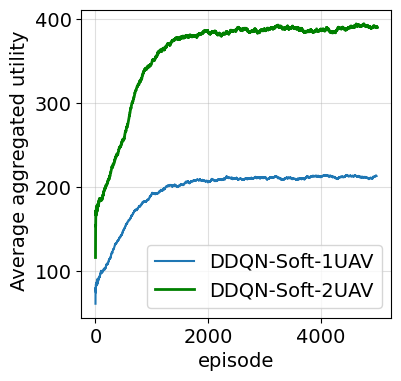}
     \vspace{-7mm}
    \caption{Training results for allocating spectrum holes to \textbf{two UAVs}.}
    \label{fig:dqn-2UAV}
\endminipage
\end{figure}

\textbf{Resource Allocation and Spectrum Access Results.}
As mentioned in Section~\ref{sec:problem2}, we use deep Q-learning methods for allocating spectrum resources to the UAVs. 
In Fig.~\ref{fig:dqn-1UAV}, we compare the training performance of three variants of Q-learning methods for allocating a sub-channel to a single UAV whenever the fusion rule detects at least a single spectrum hole. It is observed that DDQN with soft update performs slightly better and converges earlier than DDQN and vanilla-DQN. 
Next, we extend the model to allocate spectrum holes to two UAVs. In this case, we have augmented the DDQN algorithm with soft update to generate two best actions.  From the results in Fig.~\ref{fig:dqn-2UAV}, we observe that the utility performance with two SUs is slightly less than two times of the performance with a single SU. 
We further note that this paper tries to explore the possibility of integrating spectrum sensing and sharing by making use of existing RL algorithms. Though we explored Q-learning techniques, different RL algorithms can be integrated into the proposed framework.

\vspace{-3mm}




%% file: conclusion.tex
\section{Conclusion}
\label{sec:conclusion}
In this paper, we proposed a collaborative wideband spectrum sensing and sharing method for networked UAVs. 
We implemented a multi-class classifier as the spectrum sensing module to detect spectrum holes based on observed I/Q samples. To enhance the accuracy, we considered the spectrum fusion rule at the UTM server. By leveraging deep Q-learning methods, detected spectrum holes are dynamically allocated to the secondary users (i.e., UAVs). To evaluate the proposed methods, we generated a near-realistic synthetic dataset using MATLAB LTE toolbox by incorporating base-station locations in a chosen area of interest, performing ray-tracing, and emulating the primary users channel usage in terms of I/Q samples. Overall, our numerical results suggest that collaborative spectrum sensing to allocate spectrum holes to multiple SUs provide promising results.  As a future work, we are planning to conduct aerial tests to collect real-world dataset and further refine our framework. 
